\documentclass[floatfix,lengthcheck,showpacs,amssymb,amsmath,amsfonts,twocolumn,nofootinbib,longbibliography]{revtex4-2}
\usepackage{amsfonts,amsmath,units,wasysym,epsfig,graphicx,verbatim,color,subfigure,graphicx}
\usepackage{amsmath}
\usepackage{latexsym}
\usepackage{amssymb}
\usepackage{amsfonts}
\usepackage{mathtools}
\usepackage{bm}
\usepackage{color}
\usepackage{float}
\usepackage{tikz}
\usepackage{adjustbox}
\usepackage{array}
\usepackage{soul}
\usepackage{appendix}
\usepackage{physics}
\usepackage{braket}
\usepackage{xcolor}
\usepackage[none]{hyphenat}
\usepackage{lineno}
\usepackage{hyperref}
\hypersetup{colorlinks=true, citecolor=blue, urlcolor=blue, linkcolor=blue}

\newcommand{\Real}{\operatorname{Re}}

\def\be{\begin{equation}}
\def\ee{\end{equation}}
\def\bea{\begin{eqnarray}}
\def\eea{\end{eqnarray}}

\def \Msun {M_\odot}

\begin{document}

\author{\href{https://orcid.org/0000-0002-1850-4587}{Alexander H. Nitz}}
\affiliation{Department of Physics, Syracuse University, Syracuse, NY 13244, USA}

\date{\today}
\title{Robust, Rapid, and Simple Gravitational-wave Parameter Estimation}

\begin{abstract}
Rapid and robust parameter estimation of gravitational-wave sources is a key component of modern multi-messenger astronomy. We present a novel and straightforward method for rapid parameter estimation of gravitational-wave sources that uses metric-based importance sampling. The method enables robust parameter estimation of binary neutron star and binary black hole binaries and is trivially parallelized, enabling parameter estimation in seconds with modest resources. The algorithm achieves a median $35\%$ effective sampling efficiency for a population of aligned-spin neutron star binaries sources. Surprisingly, this approach is also highly efficient for analyzing the full 15-dimensional parameter space of typical binary black holes, with a population median $20\%$ efficiency achieved for a source detected primarily by the twin LIGO observatories and $9\%$ for a network of three comparable sensitivity observatories. This method can serve immediate use to improve the low-latency data products of the gravitational-wave observatory network and may be a key component of how the millions of sources observed by next-generation observatories could be analyzed. The approach can also be broadly applied for problems where an approximate likelihood metric-space can be constructed.
\end{abstract}
 \maketitle

\section{Introduction}

Gravitational waves have become a routine tool for studying the Universe and to date there are over a hundred observations of various compact-binary mergers~\cite{KAGRA:2021vkt, Nitz:2019hdf,Nitz:2021uxj, Nitz:2021zwj, Olsen:2022pin, Mehta:2023zlk, Wadekar:2023gea,Koloniari:2024kww}. The growing catalog of observations from Advanced LIGO~\cite{TheLIGOScientific:2014jea} and Virgo~\cite{TheVirgo:2014hva} is beginning to provide insights into binary formation pathways~\cite{LIGOScientific:2020ufj,Mandel:2021smh,Gerosa:2021mno,Edelman:2021fik,Zevin:2020gbd,KAGRA:2021duu}, nuclear equation of state~\cite{LIGOScientific:2018cki, Capano:2019eae}, and the search for theories beyond general relativity~\cite{LIGOScientific:2020tif, Wang:2021gqm}.

One of the most well-known observations, GW170817, was a well-localized binary neutron star merger~\cite{LIGOScientific:2017vwq}, accompanied by a gamma-ray burst~\cite{Goldstein:2017mmi,LIGOScientific:2017zic} and electromagnetic emission from a kilonova across the electromagnetic spectrum observed by dozens of observatories~\cite{LIGOScientific:2017ync}. Gravitational-wave observatories are vital to enabling multi-messenger astronomy~\cite{KAGRA:2013rdx}. To ensure reliable detection of potential neutron star binaries, several analyses have been developed which analyze data from LIGO~\cite{TheLIGOScientific:2014jea}, Virgo~\cite{TheVirgo:2014hva} and KAGRA~\cite{KAGRA}, namely PyCBC Live~\cite{Nitz:2018rgo, DalCanton:2020vpm}, GstLAL~\cite{Messick:2016aqy,Sachdev:2019vvd}, MBTA~\cite{Aubin:2020goo,Andres:2021vew}, and SPIIR~\cite{Hooper:2011rb}. These are generally able to identify a source in tens of seconds~\cite{Chaudhary:2023vec}. However, follow-up observatories need an accurate estimate of source location to know where to point and an estimate of the binary properties (e.g. component masses and spins) to determine if a given candidate is likely to have an electromagnetic counterpart~\cite{Chatterjee:2019avs}. An initial localization estimate is given by the Bayestar algorithm~\cite{Singer:2015ema} and source classification can be provided by various approaches~\cite{Kapadia:2019uut,Andres:2021vew,Villa-Ortega:2022qdo,Pankow:2015cra,Rose:2022axr}, both within a few seconds. However, the most detailed and accurate approaches rely on using Bayesian parameter estimation analyses that can take hours or days to complete~\cite{Veitch:2014wba, Berry:2014jja, Ashton:2018jfp, Biwer:2018osg}.
Robust estimates just seconds or minutes after a detection would enable telescopes to best utilize their resources and minimize the risk of missing early emission. Furthermore, as the sensitivity of current observatories improves and new observatories are added to the network~\cite{KAGRA:2013rdx}, the number of observations could increase by several orders of magnitude~\cite{Evans:2023euw}, necessitating a more efficient approach.

A number of techniques have been proposed to produce rapid parameter estimates for gravitational-wave sources. These include new methods of sampling~\cite{Tiwari:2023mzf}, speeding up the calculation of likelihoods by using reduced representations~\cite{Zackay:2018qdy,Cornish:2021lje,Smith:2016qas}, or parameter reduction by marginalization~\cite{Brown:2004vh, Singer:2015ema, Roulet:2024hwz}. Neural network enhanced sampling has recently come to the forefront as a potential method to increase the performance by orders of magnitude relative to naive MCMC or nested sampling approaches~\cite{Green:2020hst, Dax:2021tsq, Dax:2022pxd, Bhardwaj:2023xph, Williams:2021qyt, Wong:2022xvh}. These include methods that are fully likelihood-free~\cite{Bhardwaj:2023xph}, or use neural networks as a basis for proposals within a more standard algorithm~\cite{Williams:2021qyt, Wong:2022xvh}.

There are also methods that train an algorithm to directly produce a posterior estimate which is then used as the proposal for importance sampling~\cite{Dax:2022pxd}; the success of these algorithms has demonstrated the potential for amortized approaches. A disadvantage of pure neural network based approaches is that the nature of the internal model representation is not always comprehensible. This can make it difficult to robustly extend into new regimes and can result in unpredictable behavior when new and unexpected inputs are observed.

\begin{figure*}
    \centering
    \vspace*{-0.5cm}
    \includegraphics[width=1.0\textwidth]{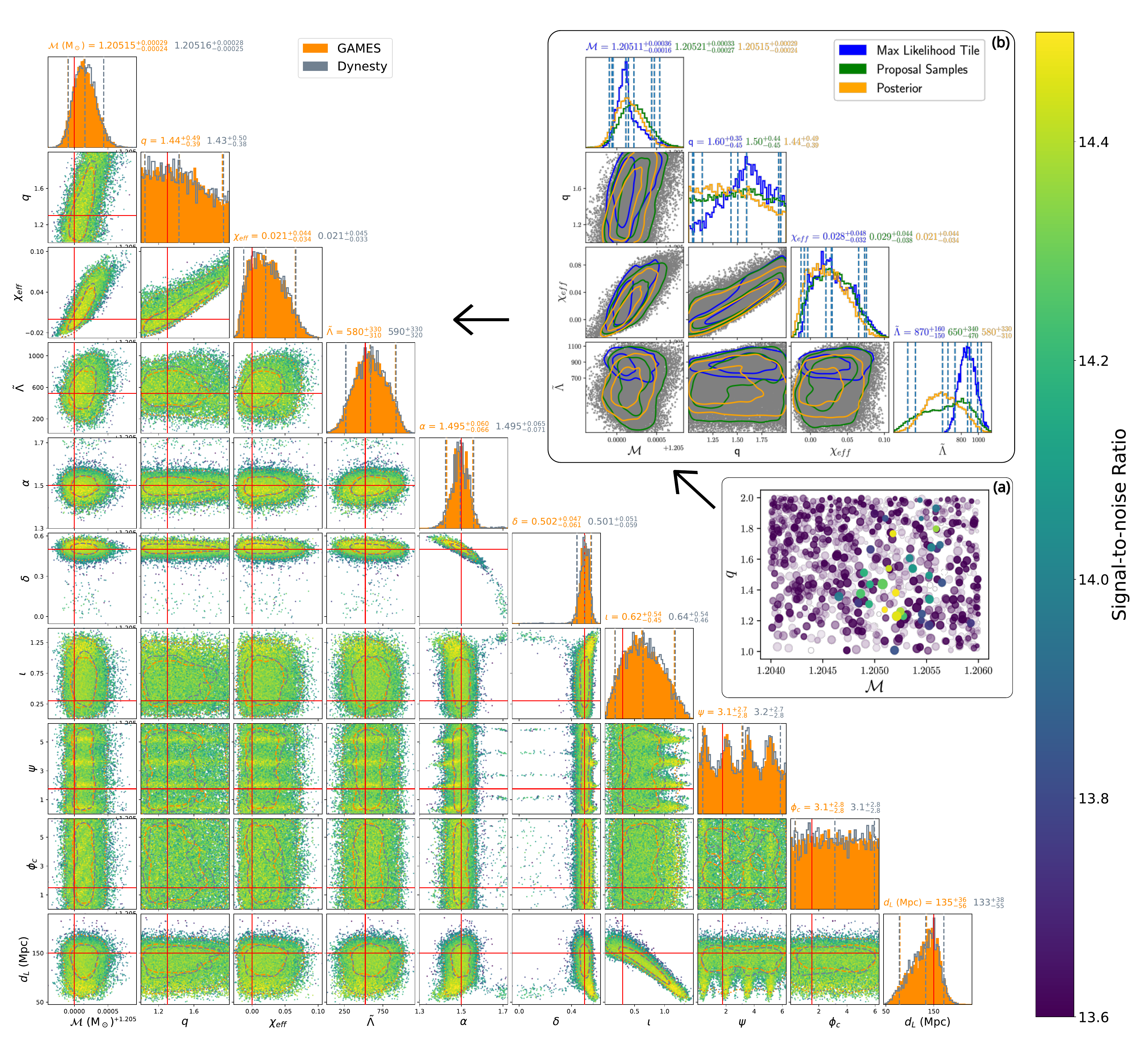}
    \caption{Key intrinsic (chirp mass $\mathcal{M}$, mass ratio q, effective spin $\chi_\textrm{eff}$~\cite{Ajith:2009bn}, effective tidal deformability $\tilde{\Lambda}$~\cite{Flanagan:2007ix}) and extrinsic parameter distributions (right ascension $\alpha$, declination $\delta$, inclination $\iota$, polarization angle $\psi$, coalescence phase $\phi_c$, luminosity distance $d_L$) for a representative simulated BNS source added to simulated noise from a three-detector network. The true parameters are shown with red lines. We compare the posterior distribution obtained from our our importance-sampling procedure (GAMES, orange) to Dynesty (gray)~\cite{speagle:2019}, an implementation of nested sampling. The sub-panels schematically show how the importance sampling procedure works. Sub-panel A shows the initial stage where the likelihood is calculated for a precalculated set of metric points. Each point is colored by the signal-to-noise; only a handful of metric points contribute to informing the final posterior. Sub-panel B shows the distribution of the prior volume contained by the largest likelihood metric tile (blue), the likelihood-weighted average of the prior used as the proposal distribution (green), and the final posterior after importance sampling (orange). Importance sampling is highly efficient because the proposal distribution is already quite close to the posterior distribution.}
    \label{fig:main}
\end{figure*}

We demonstrate a simple technique for rapidly producing gravitational wave posteriors that is based on the ability for an approximate metric of the likelihood surface to be constructed. We show that a naive brute-force tiling and exploration of this metric-space can be used as the basis for importance sampling. This approach achieves a median $35\%$ sampling efficiency for a fiducial population of binary neutron star mergers defined by their component masses, tidal deformabilities, and the spin components aligned with the orbital angular momentum; this means that an example analysis could achieve 5000 independent samples from the posterior with $\sim14,000$ likelihood calls. Furthermore, median efficiencies of $>9-20\%$ can be achieved for typical precessing binary black hole binaries. This performance is superior for typical sources to state-of-the art neural posterior methods~\cite{Dax:2022pxd} and avoids the need to train a potentially complicated latent space. Because this approach has no trained parameters and works directly with existing implementations of the gravitational-wave likelihood, it should be similarly robust to expected variations in detector noise as traditional sampling techniques based on traditional approaches that make use of markov chains or nested sampling. Given these advantages, we expect significant impacts on both low-latency and high-throughput gravitational wave science. This approach may also become a model for similar scientific problems where the signal model has a well defined metric.

\section{Metric-based Posterior Importance Sampling}

We propose a method of importance sampling to directly obtain accurate posterior samples for gravitational-wave problems. Importance sampling can allow for efficient estimation of integrals and expectations, especially in cases where direct sampling is difficult or computationally expensive. However, it is strongly affected by the choice of proposal distribution. An optimal proposal distribution is proportional to the target posterior distribution, however as that is clearly not known a priori, the aim is to construct an accurate proxy for this with a straightforward, computationally inexpensive procedure.
It is possible to efficiently create an accurate proposal distribution for gravitational-wave problems because the likelihood can be defined in terms of an inner product whose salient features can be exhaustively mapped independent of any specific observation. This allows for an approximate metric-space to be constructed and for the prior volume to be efficiently tiled. A proposal distribution can be constructed by simply drawing from the prior volume associated with each tile in proportion to the approximate posterior probability contained within each tile. A similar approach would likely apply to any problem of this nature.

\subsection{Gravitational-wave Likelihood}
\label{sec:likelihood}

We'll start by motivating why a metric space can be constructed. The likelihood for gravitational-wave data analysis can be expressed in terms of the inner product

\begin{equation}
    (a|b) = 4\Real\int \frac{a(f)^* b(f)}{S_{n}(f)}  df
\end{equation}

in the case of frequency-domain functions a, b and the noise power spectral density $S_n$. The full likelihood can be written as

\begin{equation}
ln\mathcal{L} = {-\frac{1}{2}{(d-h|d-h)}} = (d|h) - \frac{1}{2}(h|h) - \frac{1}{2}(d|d)
\end{equation}

where d is the data and h is the signal proposed to be in the data. This form naturally results from the assumption of wide-sense stationary colored Gaussian noise at the time of a potential signal. This is the standard assumption in gravitational-wave astronomy, though deviations from this have been explored~\cite{Veitch:2014wba}. Since the data is the sum of the noise n and an actual source signal s, the ratio of the additive signal to noise-only hypothesis can be written as

\begin{equation}
    ln(\mathcal{L}/\mathcal{L}_n) = (n|h) + (s|h) - \frac{1}{2} (h|h).
\end{equation}

If we focus on factors related to the phase evolution of a signal and are independent of its amplitude, we can maximize the likelihood to give

 \begin{equation}
    ln(\mathcal{L}/\mathcal{L}_n)_{maxA} = (n|h) + \frac{(s|h)^2}{2(h|h)} = (n|h) + {\frac{1}{2}\mathcal{O}(h|s)^2 (s|s)}.
\end{equation}

where $\mathcal{O}$ is the normalized overlap between the source and proposed signal. The non-trivial structure of this likelihood as a function of the proposed template $h(\theta)$ arises from the behavior of $\mathcal{O}(h|s)$. This can be viewed as the inner product of the space defined on the gravitational waveform manifold and can be fully explored independently of specific observational data. Our algorithm recasts parameter estimation in terms of proposals defined naturally on the metric space defined by this inner product.

 Constructing a set of points within the manifold defined by this metric that provides an efficient packing is a well-known problem that has been explored as `template bank generation` in the context of searches for gravitational-wave sources~\cite{Owen:1995tm}. There exist several algorithms for tiling this space such that coverage is achieved with a minimum metric distance maintained between points in the space~\cite{Ajith:2012mn, Brown:2012qf, Kacanja:2024pjh}. Because the likelihood can be expressed entirely in terms of this inner product, a tiling of the metric space is also an approximate tiling of the likelihood space. Notably, we don't require the metric to exactly match that of the likelihood itself, but only be consistent for small patches. For a small region, terms in the likelihood such as $(n|h)$ can be treated as nearly constant with small changes bounded by a triangle inequality.

Once the metric space is tiled, we determine the portion of the prior volume closest to each tiling point. To generate the posterior, we first calculate the posterior probability for each tiling point; there are typically few tiling points that need to be calculated $\mathcal{O}(10^{3-4})$ in practice when using a information from an initial search identification. The proposal distribution is then the portion of prior volume associated with each tile weighted by the posterior probability of its associated tiling point. In Fig~\ref{fig:main}, we show schematic picture of this process starting from the template bank and likelihood with prior regions already encapsulated.

\begin{figure*}
    \centering
    \vspace*{-0.5cm}
    \includegraphics[width=0.85\textwidth]{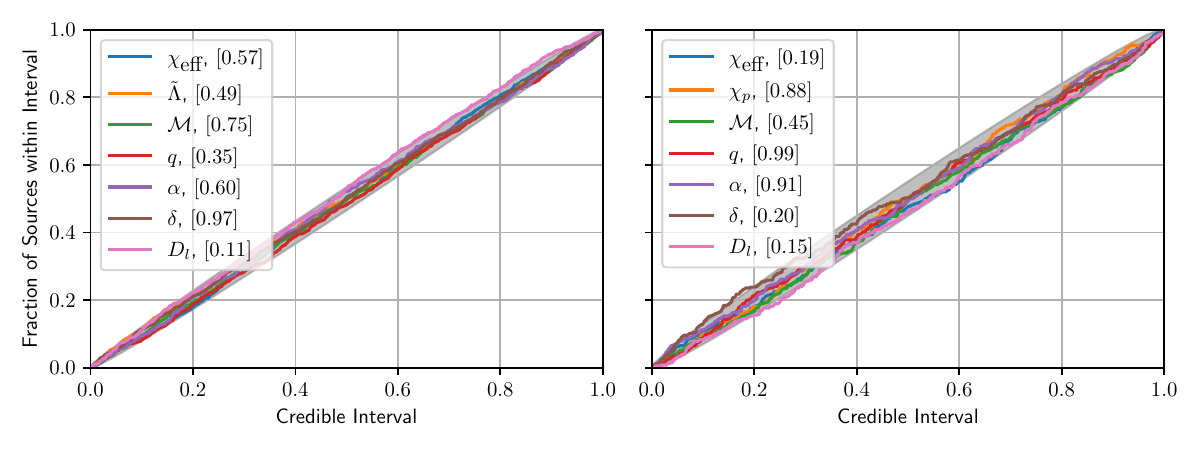}
    \caption{The fraction of simulated sources with true values within a credible interval for our BNS (left) and BBH population (right) along with binomial pointwise 2-$\sigma$ confidence bands (gray). For each parameter we perform a KS test (shown in legend); the p-values are consistent with the algorithm producing self-consistent posteriors. The combined p-value is 0.70 and 0.57, for the BNS and BBH simulations, respectively.} Key parameter combinations are shown ($\mathcal{M}$, $q$, $\mathcal{\chi_\textrm{eff}}$, $\chi_p$~\cite{Schmidt:2014iyl}, $\tilde{\Lambda}$~\cite{Flanagan:2007ix}) along with the extrinsic parameters right ascension $\alpha$, declination $\delta$ and luminosity distance $D_l$.
    \label{fig:pp}
\end{figure*}

\begin{figure}
    \centering
    \vspace*{-0.5cm}
    \includegraphics[width=0.5\textwidth]{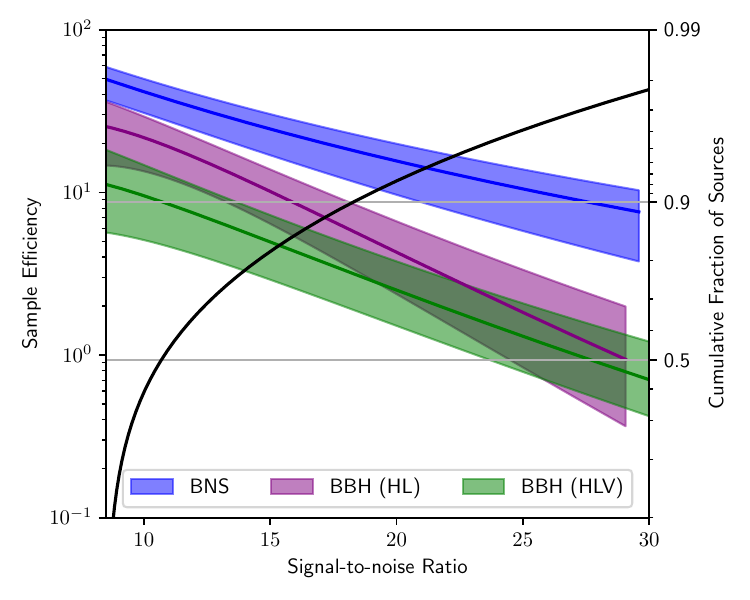}
    \caption{The sample efficiency as a function of signal-to-noise ratio for aligned spin BNS (blue), three-detector BBH (green), and two-detector BBH sources (purple). In each case, the efficiency is calculated over the overage of 80,000 likelihood calls. The cumulative number of detectable sources is shown in black.
    }
    \label{fig:scale}
\end{figure}

\section{Application to Binary Neutron Stars and Binary Black Holes}
\label{sec:application}
We examine two specific cases of interest. The first case is that of binary neutron stars, which we'll characterize by their masses, aligned spin components, and tidal deformabilities. Second, we will examine binary black holes of typical mass ($\mathcal{M} \sim ~33\Msun$) with the full parameter space of spin orientations, allowing for precession and non-negligible higher-order modes. This covers the most typical cases for the current observed population of sources. We assume natural priors for extrinsic parameters (e.g. sky location, orientation) and allow for spin magnitudes up to 0.4 and 0.85 for the BNS and BBH cases, respectively; these choices span the observed population and are motivated by the spin of observed millisecond pulsars~\cite{Hessels:2006ze} and the well-calibrated regions of existing models, respectively~\cite{Pratten:2020ceb,Varma:2019csw}. To model the gravitational-wave signal, we use the IMRPhenomD\_NRTidal (BNS)~\cite{Khan:2015jqa, Dietrich:2017aum} and IMRPhenomXPHM (BBH)~\cite{Pratten:2020ceb} waveform models.

We use our algorithm to estimate the intrinsic parameters of the source in both cases, while the other parameters are marginalized over using SNR-informed importance sampling~\cite{Roulet:2024hwz} or analytic formulas~\cite{Brown:2004vh, Singer:2015ema}. In the case of precessing BBH signals, we also choose to sample over inclination explicitly rather than marginalizing over the parameter; we also numerically marginalize the orbital phase in this case. We use this marginalized likelihood for the weights of the metric-space importance sampling.

We construct a tiling of the metric space describing the similarity between signal waveforms using the stochastic algorithm introduced in ~\cite{Kacanja:2024pjh}; the metric distance between tiles is chosen so that they have a mismatch of $1\%$ ($1.5\%$) for BNS (BBH) sources. Each tile is associated with a portion of the prior volume. The prior is chosen as discussed above and is independent of the tiling; only the fraction of the total prior volume associated with each tile changes with different choices of prior. This mapping between the prior volume and the tiled metric space is constructed by explicit numerical waveform comparison of a large sets of model waveforms $O(10^8)$ drawn according to the prior. The tiling and prior mapping is done separately for different slices of chirp mass; in a production analysis a slice containing $O(10^3-5)$ templates can be identified based on preliminary information from a detection analysis. For specific parameter cases, a known analytic metric could potentially be used to more quickly map the prior volume. We simplify the tiling process by taking advantage of the fact that the intrinsic and extrinsic parameters partially decouple~\cite{Singer:2015ema}; this is a weaker approximation for highly precessing signals which does reduce the overall efficiency. In the case of aligned spin signals, the association between metric point and prior volume is done using the simple waveform overlap, whereas for precessing systems we approximate this as the minimum of the overlap between the metric point and the individual plus and cross gravitational waveform polarizations; for aligned spin systems the polarizations differ only by an overall phase shift.

The tiling is only weakly sensitive to expected changes in the power spectral density $s_n(f)$ as has been studied in the context of searches~\cite{Usman:2015kfa}. We would expect that the space would only need to be remapped when there are significant changes such as between observing runs. However, even in these cases, the primary impact would be to stretch and squeeze the underlying space, potentially decreasing the effective metric density of templates; the primary impact would be a reduction in sampling efficiency which can be easily monitored through significant instrumental sensitivity changes. The method can also be straightforwardly extended to incorporate calibration and waveform uncertainties either directly through the metric space of the waveform model or by using a calibration-marginalized likelihood~\cite{Essick:2022vzl}.

\subsection{Validity and Efficiency}
\label{sec:validity}
To test the self-consistency of our method, we simulate a population of sources added to simulated observatory noise. We consider a network of three observatories with sensitivity consistent with Advanced LIGO's third observing run. For each source, we follow our method and estimate the parameters of each source. In Fig.~\ref{fig:pp} we demonstrate that the resulting estimates are consistent with unbiased probability distributions. A handful of individual sources are also directly compared against the Dynesty sampler as configured in past analyses~\cite{Nitz:2021zwj}. In each case, the results are indistinguishable. One such example is shown in Fig.~\ref{fig:main}. To test the robustness of the approach in cases where the noise curve may have deviated from what it was constructed for, we apply the same pre-computed metric tiling used for simulated O3 data to O4 simulated data; this also produces unbiased posterior estimates.

To assess the efficiency of the algorithm, we calculate the effective sample size for each simulated source relative to the number of likelihood evaluations. Fig.~\ref{fig:scale} shows how the sample efficiency scales as a function of signal SNR. For the BNS case, high efficiency $>10\%$ is achieved for signals up to SNR of 30 and the median of the population is $35\%$. While it is perhaps not surprising that for a simple enough parameter space this approach may be efficient, we find that it can also handle the 15 dimensional parameter space of typical BBH mergers including precession effects and the inclusion of higher order modes. Because the metric mapping is further approximated in comparison to the aligned spin BNS case, we find a reduction in the sampling efficiency, however, we note that for the majority of sources it is still $>10\%$ which is competitive with the state-of-art neural importance sampling methods~\cite{Dax:2022pxd}.

Given the naive implementation of this method tested here, we would expect the efficiency to drop with increasing SNR as is observed. The drop in efficiency occurs for two reasons: (1) the fixed metric tiling distance means that as the SNR of a signal increases, the coarseness of the tiling is increasingly apparent in the proposal distribution and (2) the intrinsic dimensional of the metric space increases as more subtle parameter effects become measurable. A more sophisticated treatment of the metric space in this case, would very likely result in further efficiency improvements by recording the details of the local metric around each of the prior points and incorporating that into an iterative procedure to update the proposal distribution.

\section{Discussion}

We have demonstrated a novel approach to rapid and robust gravitational-wave parameter estimation using metric-based importance sampling. The algorithm is available through the PyCBC toolkit~\cite{pycbc-github} and a corresponding data release provides a working example~\cite{github}. This method is effective for the majority of typical gravitational-wave sources, including binary neutron star (BNS) and typical binary black hole (BBH) systems. Our approach, which tiles the parameter space using an approximate metric, achieves high sampling efficiencies— a population median of 35\% for BNS and 9–20\% for BBH sources. A key advantage of this method is its inherent interpretability while at the same time achieving state-of-the-art efficiency comparable to machine learning approaches~\cite{Dax:2022pxd}; While machine learning models can be opaque, our algorithm's foundation in first principles allows its behavior and potential biases to be fully characterized.

 Although the method handles most sources efficiently, it does not currently eliminate the need for other parameter estimation algorithms. It may prove more difficult to extend to highly precessing NSBH or lower mass BBH mergers where the intrinsic parameter space size is significantly higher. Furthermore, the current implementation is not optimal for extremely loud signals, where precision is crucial, or for cases where the waveform model is not well-understood in advance. In situations where a full tiling of the parameter space is impractical, more general algorithms will still be necessary. These edge cases will likely be of high scientific interest and will require further development of this approach and complementary techniques. However, our method offers a clear advantage for the vast majority of sources~\cite{Nitz:2021zwj,KAGRA:2021vkt}, significantly reducing computational costs and time.

We have not yet integrated this method into a full low-latency pipeline for real-time gravitational-wave parameter estimation, but the potential for such an implementation is clear. After non-trivial optimization of data ingestion, the main time constraint would be likelihood evaluations. For example, achieving 10,000 effective samples for BNS sources could be completed in O(10) seconds on a machine with O(10) cores with the tested likelihood implementation. This would drastically improve the ability to deliver fast, reliable source estimates to multi-messenger partners, enhancing the chances of capturing early electromagnetic counterparts. Since our improvements are to sampling efficiency, parallel developments in the performance of likelihood calculations can be directly leveraged in a production analysis.

\begin{acknowledgments}
We would like to acknowledge Collin D. Capano for fruitful discussions. AHN acknowledges support from NSF grant PHY-2309240. AHN acknowledge the support from Syracuse University for providing the computational resources through the OrangeGrid High Throughput Computing (HTC) cluster.

This research has made use of data or software obtained from the Gravitational Wave Open Science Center (gwosc.org), a service of the LIGO Scientific Collaboration, the Virgo Collaboration, and KAGRA. This material is based upon work supported by NSF's LIGO Laboratory which is a major facility fully funded by the National Science Foundation, as well as the Science and Technology Facilities Council (STFC) of the United Kingdom, the Max-Planck-Society (MPS), and the State of Niedersachsen/Germany for support of the construction of Advanced LIGO and construction and operation of the GEO600 detector. Additional support for Advanced LIGO was provided by the Australian Research Council. Virgo is funded, through the European Gravitational Observatory (EGO), by the French Centre National de Recherche Scientifique (CNRS), the Italian Istituto Nazionale di Fisica Nucleare (INFN) and the Dutch Nikhef, with contributions by institutions from Belgium, Germany, Greece, Hungary, Ireland, Japan, Monaco, Poland, Portugal, Spain. KAGRA is supported by Ministry of Education, Culture, Sports, Science and Technology (MEXT), Japan Society for the Promotion of Science (JSPS) in Japan; National Research Foundation (NRF) and Ministry of Science and ICT (MSIT) in Korea; Academia Sinica (AS) and National Science and Technology Council (NSTC) in Taiwan.
\end{acknowledgments}
\clearpage
\bibliography{references}

\clearpage
\appendix

\section{Sampling Procedure Demonstration}

\begin{figure}
    \centering
    \vspace*{-0.5cm}
    \includegraphics[width=0.5\textwidth]{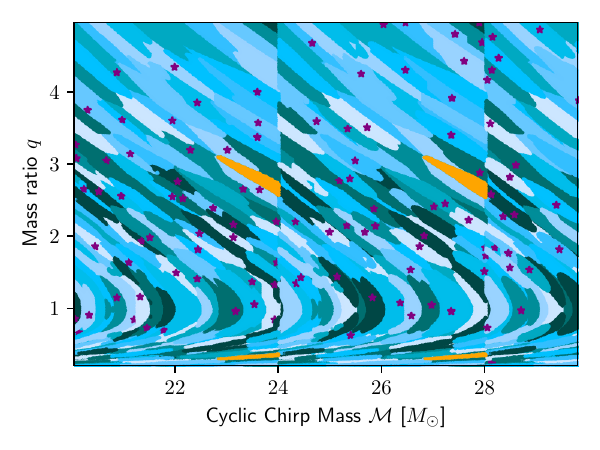}
    \caption{The locations of templates that span our example space (purple stars) and the associated tiles (shades of blue) that subdivide the prior volume as a function of mass ratio $q$ and our modified "cyclic" chirp mass $\mathcal{M}$. Each tile is defined by the locus of prior samples that is in closest proximity to a template point. One example tile is shown in orange. Notably, this tile can be arbitrarily multi-modal in a physical parameterization such as demonstrated here where there are four modes in the physical coordinate system. The strength of the algorithm partially results from the fact that the sampling does not depend on the physical coordinates and so multi-modality does not have an impact.  
    }
    \label{fig:af0}
\end{figure}

\begin{figure*}
    \centering
    \vspace*{-0.5cm}
    \includegraphics[width=1\textwidth]{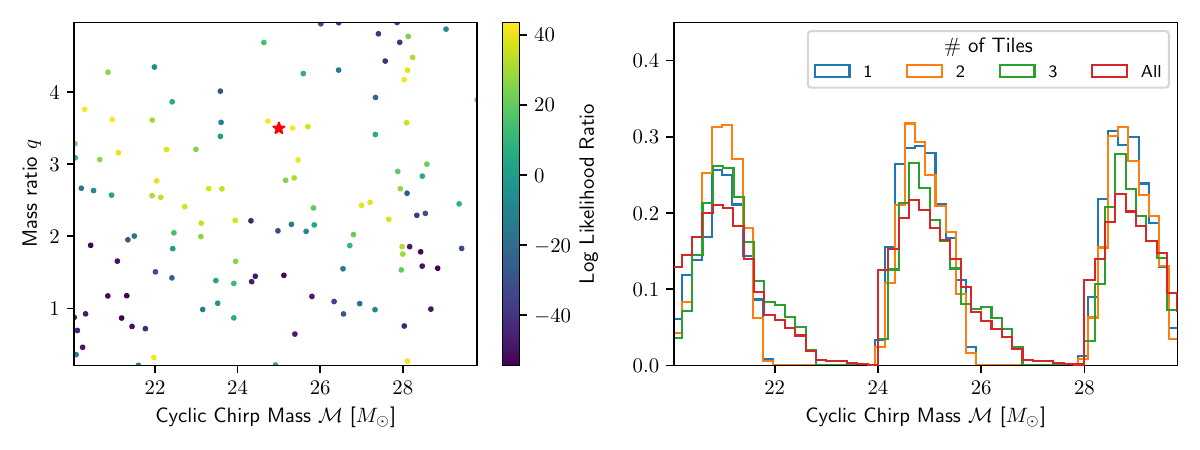}
    \caption{Left: the locations of templates that span our example space as a function of mass ratio $q$ and cyclic chirp mass $\mathcal{M}$ colored by the likelihood ratio for the parameters associated with each template. The posterior probability associated with each tile forms the basis for the proposal sampling distribution. Right: the posterior probability weighted distribution of the value of cyclic chirp mass for the prior samples that make up the most significant one (blue), two (orange), or three (green) tiles, along with the final proposal distribution formed from all the tiles (red). The essential shape of the distribution, including the possibility of multiple modes in the physical parameter space, is recovered with only a small number of tiles.
    }
    \label{fig:af1}
\end{figure*}

\begin{figure*}
    \centering
    \vspace*{-0.5cm}
    \includegraphics[width=1.0\textwidth]{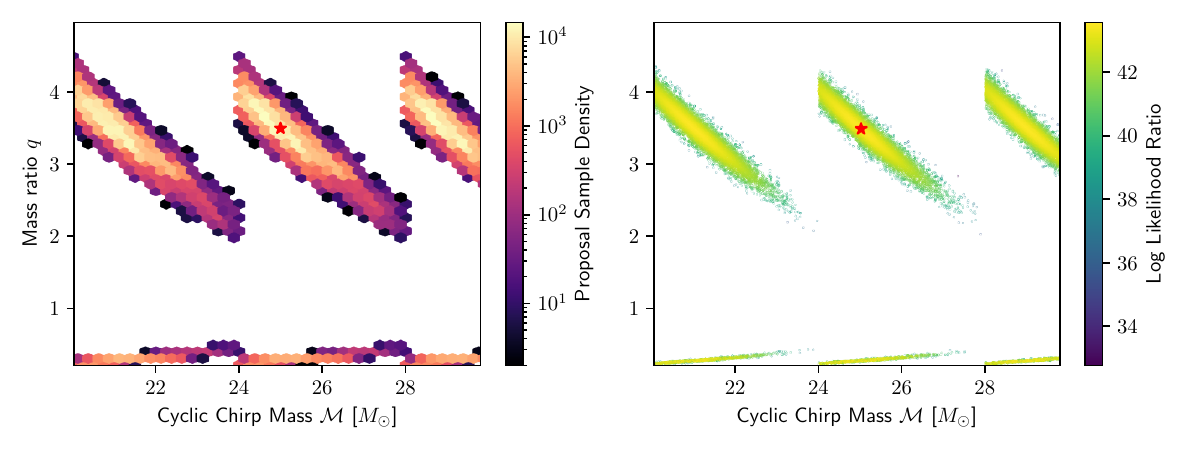}
    \caption{Left: the proposal distribution density as a function of mass ratio $q$ and cyclic chirp mass $\mathcal{M}$. Right: samples from the posterior distribution produced by our metric-based sampling algorithm colored by their log likelihood ratio. The true injected parameters are shown with a red star. We see that the full multi-modality of the space is recovered. The effective sampling efficiency for this simplified problem is $65\%$ confirming that the proposal distribution is already a close match to the final posterior distribution.
    }
    \label{fig:af3}
\end{figure*}

\begin{figure*}
    \centering
    \vspace*{-0.5cm}
    \includegraphics[width=1\textwidth]{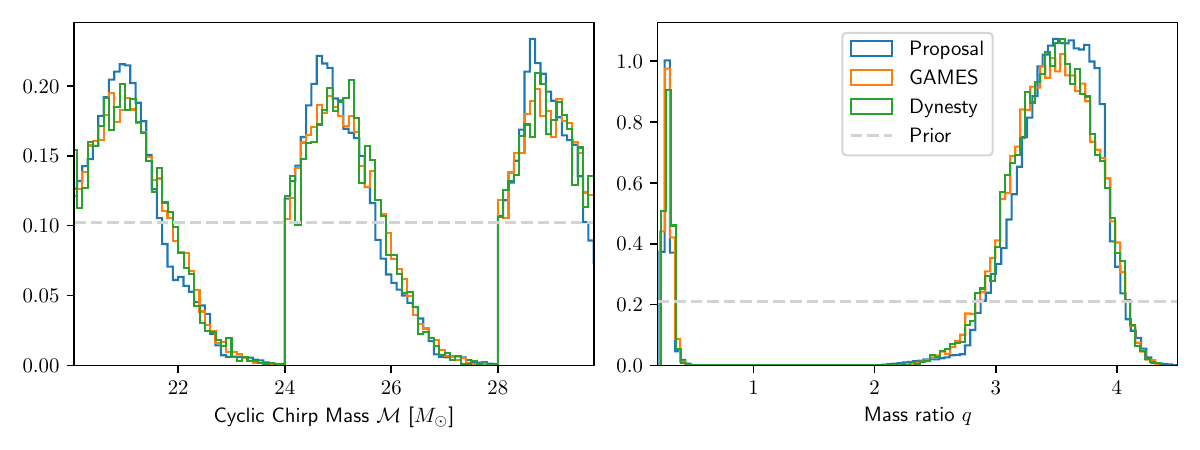}
    \caption{The posterior distribution for cyclic chirp mass $\mathcal{M}$ (left) and mass ratio $q$ (right) produced by our metric-based sampler (GAMES, orange), a nested sampling implementation (Dynesty, green), and the initial proposal distribution used in our method (blue). We use a uniform prior distribution in these parameters (gray dashed lines). Both the nested sampling algorithm and our metric-based sampling approach give consistent results. 
    }
    \label{fig:af2}
\end{figure*}

To demonstrate the procedure of tiling and mapping a parameter space, we'll start with a simplified example. We consider a binary black hole merger whose instrinsic parameters are characterized by just two parameters, its chirp mass $\mathcal{M}$, and its mass ratio $q$. We create a simulated source whose true parameters are $\mathcal{M}=25$, $q=3.5$. To highlight one of the unique advantages in this approach to naturally handle multi-modal spaces, we will sample over $q < 1$ and $q > 1$; this produces two modes because the signal is unchanged when switching the label of which is the primary and secondary mass. We further complicate the space by defining a new parameter "cyclic chirp mass" which is defined as $\mathcal{M}_{\textrm{cyclic}} = \textrm{mod}(\mathcal{M}, 4) + 24$.  

Once the target space is defined, there is a precomputation step which maps the signal model space. This can be split into two steps: (1) the creation of a template bank which spans the target parameter space and (2) the calculation of which parts of the total prior volume are closest to each template. These two steps create tiles of the original prior volume that span the entire space. 

\subsection{Template bank generation}
As discussed in Secs~\Ref{sec:likelihood} and~\Ref{sec:application}, the first stage of understanding the parameter space is to tile it efficiently by constructing a template bank in the parameters of interest. We have found that existing stochastic template bank generation~\cite{Kacanja:2024pjh} is sufficient for this task, however, we note that alternative approaches may also work in practice. A template bank is a discrete set of points in the parameter space which typically satisify a minimal match criteria, whereby for a signal anywhere in the parameter space, there will be a corresponding template that has as a signal agreement (match) at least as high as some threshold. This sthreshold isoften chosen to be $\sim0.965$ for searches. For this demonstration we choose a high value of 0.996 which yields $\sim 100$ templates. The location of these templates is shown as purple stars in Fig.~\ref{fig:af0}. Since the template bank is used to define the tiles of our overall parameter space, there is in general motivation to use as high a minimal match as possible; however this is balanced by the practicality in doing so for the space under consideration. The efficiency of the sampling procedure depends on the similarity between the proposal distribution and the target distribution. As the SNR of a signal increases, more of the posterior is concentrated in a smaller number of \textit{tiles} hence the sampling efficiency should decrease for a tiled parameter space using a constant minimal match. As expected, we see this decrease in Fig.~\ref{fig:scale}. 

 Since the metric defined by the match inner product also closely correponds to a metric on the likelihood surface itself, a uniform tiling of the space using the match criteria is also a tiling of the likelihood surface. This means that although the template bank density may vary as a function of the physical parameters, they are uniformly distributed in the intrinsic space of waveform similarity. As an example, the high mass region of a template bank has comparatively few templates relative to the lower mass end. While in this physical parameter space, the density of template is non-uniform, in the space of signal waveform similarity it is uniform. This is consistent with the fact that higher mass sources at a fixed SNR have have higher relative uncertainties on their parameters.

\subsection{Tiling the Prior Volume}

 The second step of the precomputation procedure is to map the prior volume into a set of tiles. Each tile is defined by the locus of samples from the prior volume that are closest to a particular template point, where distance is defined by the straightforward waveform similarity (match). The effect of this procedure to tile the space is shown in Fig.~\ref{fig:af0}. 

\subsection{Generation of the posterior distribution}



The posterior distribution in our method is produced using importance sampling. The posterior probability distribution is sampled using an easier to sample proposal distribution $g(\theta_i)$. Samples are drawn from the proposal distribution, their posterior probability is calculated and is poportional to $L(D|\theta_i)\pi(\theta_i)$ where $L$ and $\pi$ are the likelihood and prior, respectively. The samples are finally re-weighted based on their drawing probability to recover samples from the posterior distribution. In the trivial case where the proposal distribution matches the posterior distribution, the weighting factors would be unity and the proposal samples would already be equally-weighted posterior samples, however, in general the weighting factor can be expressed as 

\begin{equation}
    w'_i = \frac{L(D \mid \theta_i)\pi(\theta_i)}{g(\theta_i)}.
\end{equation}

A key metric often used in importance sampling techniques is the \textit{effective sampling efficiency}. This is a measure of the fraction of independent samples of the posterior distrubution that are produced relative to the number of proposal samples that had to be evaluated. A common approximation for the effective number of independent samples used throughout the literature~\cite{Doucet2001,liu2008,10.1111/j.1541-0420.2011.01553.x} is 

\begin{equation}
 \text{ESS} = \frac{\left(\sum_{i=1}^{N} w'_i\right)^2}{\sum_{i=1}^{N} (w'_i)^2}.
\end{equation}

In the case where the proposal distribution and the posterior distribution are identical, the effective sample size becomes the same as the number of proposal samples and the effective sampling efficiency would be $100\%$.

To produce the proposal distribution for our metric-based method, we start by calculating the posterior probability at the location of each template of our template bank. In the left side of Fig.~\ref{fig:af1}, we see the templates of our demonstration example along with each associated likelihood value. We generate proposal samples from each tile in proportion to the size of the prior volume encapsulated in each tile and the likelihood calculated for the associated template. We can see from the right side of Fig.~\ref{fig:af1} that even taking just the few most important tiles produces a distribution that reflects the complexity of the space. 

Finally, the posterior probability is calculated for each proposal sample and the final posterior is formed by re-weighting these samples. The proposal distribution density is shown in Fig.~\ref{fig:af3} along with the resulting posterior distribution. In Fig.~\ref{fig:af2} we compare our final posterior distribution to that obtained by the Dynesty algorithm; we find consistent results. This two-dimensional demonstration case is highly simplified from the full gravitational-wave problem, however, even in this case, we find that our metric-based sampling approach is $10\times$ more efficient than Dynesty, resulting in an effective sampling efficiency of $65\%$ vs $6\%$. 


\section{Examples sources from the Gravitational-wave Catalog}

We examine a few representative examples from the gravitational-wave catalog using the new metric-based importance sampling method. In Section ~\ref{sec:validity} we analyzed a population of simulated sources injected into Gaussian noise. We demonstrate that the key findings of these simulations extend to realistic sources, both in terms of the efficiency of the sampling, but the reproduction of results obtained with standard sampling methods.

The same astrophysical priors are assumed as described in section ~\ref{sec:application}. For each signal, we generate a precomputed tiling of the parameter space and mapping to the prior volume using the average noise curve of the observing run containing the event. We compare the posterior results for each observation to those obtained with the Dynesty sampler, configured similarly to the production configuration used for the 4-OGC analysis~\cite{Nitz:2021zwj}.

\begin{figure*}
    \centering
    \vspace*{-0.5cm}
    \includegraphics[width=1.0\textwidth]{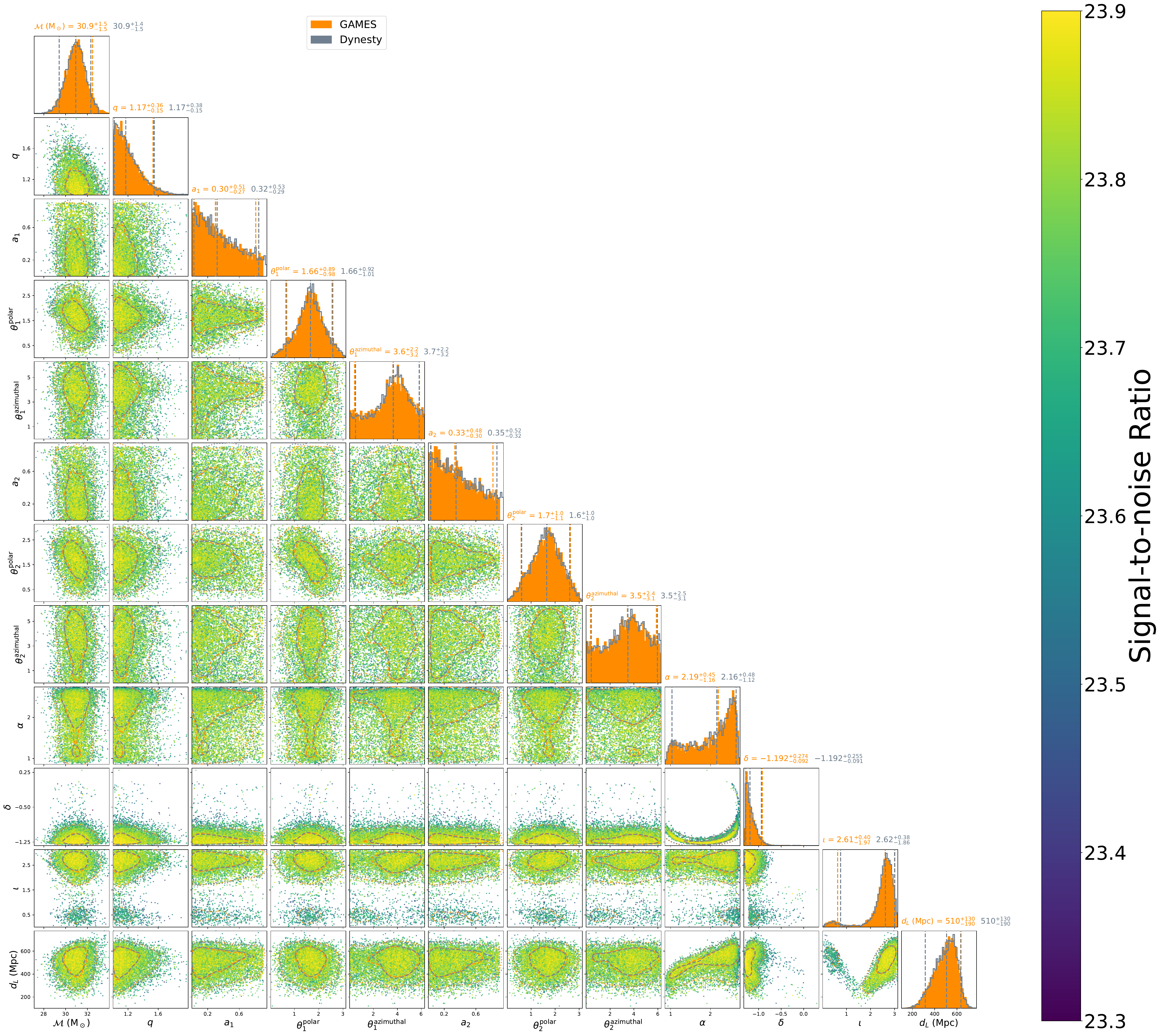}
    \caption{The posterior distribution of GW150914 for our metric-based sampling method (Games, orange) and an implementation of nested sampling (Dynesty, gray). Contours are shown for the $90\%$ credible region. The key intrinsic parameters shown are the detector-frame chirp mass $\mathcal{M}$, mass ratio $q$, and spins of each component object defined by their respective spin vector amplitude $a_{1,2}$, polar angle $\theta^{\textrm{polar}}_{1,2}$, and azimuthal angle $\theta^{\textrm{azimuthal}}_{1,2}$. In addition, we show the extrinsic parameters right ascension $\alpha$, declination $\delta$, inclination $\iota$, and luminosity distance $d_L$. 
    }
        \label{fig:150914}
\end{figure*}

\begin{figure*}
    \centering
    \vspace*{-0.5cm}
    \includegraphics[width=1.0\textwidth]{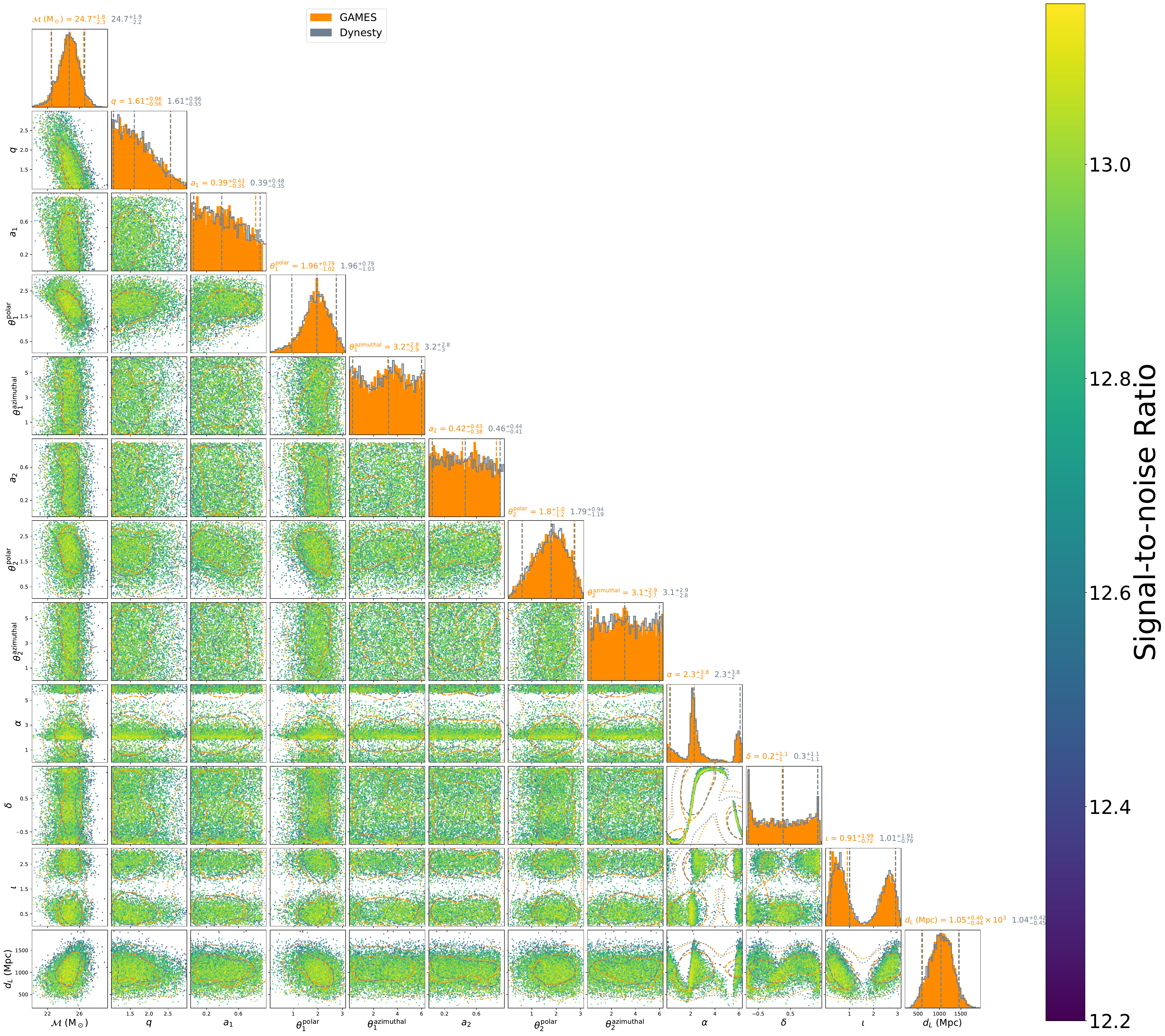}
    \caption{The posterior distribution of GW170104 for our metric-based sampling method (Games, orange) and an implementation of nested sampling (Dynesty, gray). Contours are shown for the $90\%$ credible region. The key intrinsic parameters shown are the detector-frame chirp mass $\mathcal{M}$, mass ratio $q$, and spins of each component object defined by their respective spin vector amplitude $a_{1,2}$, polar angle $\theta^{\textrm{polar}}_{1,2}$, and azimuthal angle $\theta^{\textrm{azimuthal}}_{1,2}$. In addition, we show the extrinsic parameters right ascension $\alpha$, declination $\delta$, inclination $\iota$, and luminosity distance $d_L$.  }
    \label{fig:170104}
\end{figure*}

\begin{figure*}

    \centering
    \vspace*{-0.5cm}
    \includegraphics[width=1.0\textwidth]{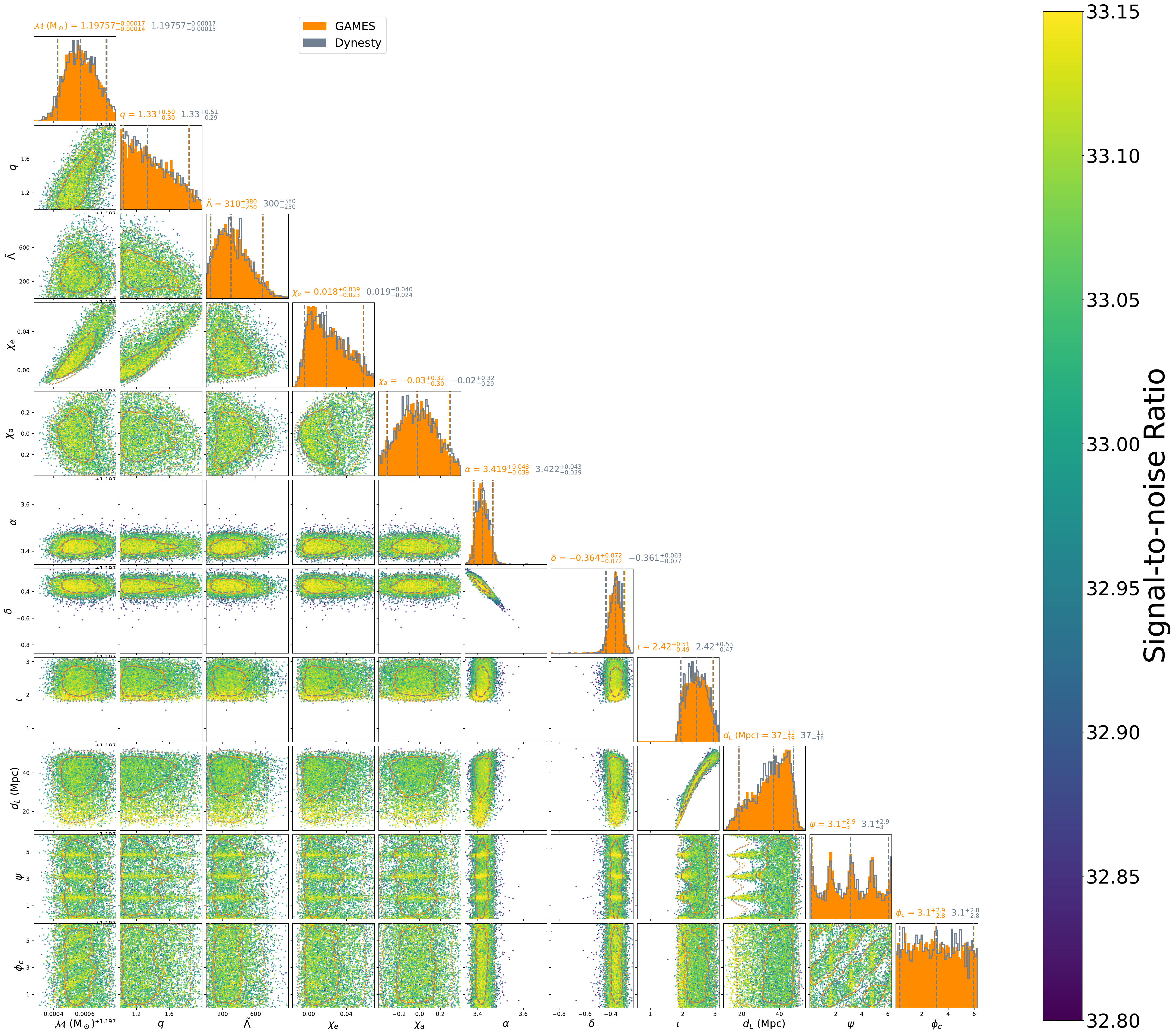}
    \caption{The posterior distribution of GW170817 for our metric-based sampling method (Games, orange) and an implementation of nested sampling (Dynesty, gray). Contours are shown for the $90\%$ credible region. The key intrinsic parameters shown are the detector-frame chirp mass $\mathcal{M}$, mass ratio $q$, combined tidal deformability parameter $\tilde{\Lambda}$, and spins of each component object defined by the combined effective spin parameter $\chi_e$ and the asymmetric spin combination $\chi_a$. In addition, we show the extrinsic parameters right ascension $\alpha$, declination $\delta$, inclination $\iota$, polarization angle $\psi$, orbital phase $\phi_c$, and luminosity distance $d_L$.}
    \label{fig:170817}
\end{figure*}

The posterior distributions for the two binary black hole mergers GW150914~\cite{LIGOScientific:2016aoc} and GW170104~\cite{LIGOScientific:2017bnn} along with the neutron star binary merger GW170817~\cite{LIGOScientific:2017vwq} are shown in Figs.~\ref{fig:150914},~\ref{fig:170104}, and~\ref{fig:170817}, respectively. In each case, the distributions obtained from our metric-based sampling algorithm and Dynesty are in agreement. The sampling efficiency of each is also consistent with that observed in Fig.~\ref{fig:scale} for their respective source type and signal-to-noise ratio, with an effective sampling efficiency of $3.5\%$, $13\%$, and $3\%$, respectively. This is a $30-100\times$ improvement in sampling efficiency relative to the compared Dynesty configuration.

\section{Expected Performance}

\begin{center}
\begin{tabular}{||c |c |c |c||} 
 \hline
Signal & SNR & CPU Cores & Time [s] \\ [0.5ex] 
 \hline\hline
BNS & 10 & 32 & 8 \\ 
BNS & 20 & 32 & 18 \\ 
BNS & 30 & 32 & 40 \\ 
BNS & 30 & 128 & 10 \\
BBH (HL) & 10 & 32 & 60 \\ 
BBH (HLV) & 10 & 32 & 120 \\ 
BBH (HL) & 20 & 32 & 300 \\ 
BBH (HLV) & 20 & 32 & 500 \\ 
 \hline
\end{tabular}
\end{center}
\label{table:cost}

The expected computational performance characteristics for a few key computing scenarios is shown in Table~\ref{table:cost} assuming no further improvements are made to the sampling approach or the likelihood implementations. In each case, we use our BNS and BBH simulated examples as the reference and assume a target of 10,000 independent samples. Furthermore, we assume that the cost is dominated by the likelihood evaluations. Parallel work into improving the computational efficiency of likelihood evaluation, or simple implementation on GPU hardware would significantly improve these metrics. The reference CPU core performance is that of an AMD EPYC Rome CPU where each likelihood calculation (including the cost of extrinsic parameter marginalization) takes 10 and 40 ms for the BNS and BBH case, respectively.

\end{document}